\documentclass[aps,twocolumn,superscriptaddress]{revtex4}
\usepackage{graphicx}% Include figure files
\usepackage{slashbox}

\begin{document}

\title{Dynamics of a two-species Bose-Einstein condensate in a double well}

\author{B. Sun and M. S. Pindzola}

\affiliation{Department of Physics, Auburn University, Auburn, AL
36849, USA}

\begin{abstract}
We study the dynamics of a two-species Bose-Einstein condensate in
a double well. Such a system is characterized by the intra-species
and inter-species $s$-wave scattering as well as the Josephson
tunnelling between the two wells and the population transfer
between the two species. We investigate the dynamics for some
interesting regimes and present numerical results to support our
conclusions. In the case of vanishing intra-species scattering
lengths and a weak inter-species scattering length, we find
collapses and revivals in the population dynamics. A possible
experimental implementation of our proposal is briefly discussed.
\end{abstract}


\maketitle

\section{introduction}

Soon after the realization of scalar Bose-Einstein condensates
(BECs) in laboratories, there has been great interest in exploring
two species BECs both experimentally
\cite{wieman,cornell0,cornell,ketterle0,ketterle,cornell2,
holland,ingu0,ingu1,grimm,ingu2,wieman2} and theoretically
\cite{ho,eberly,bigelow,bigelow2,graham,ricardo,carr,bohn,ricardo2}.
Such a two species BEC provides an ideal platform for the study of
more intriguing phenomena. It can help us understand some basic
problems in ultralow temperature physics like interpenetrating
$^3$He-$^4$He mixtures. The two species can be two alkali like
$^{23}$Na-$^{87}$Rb or two isotopes like $^{85}$Rb-$^{87}$Rb. It
can also be two hyperfine states of the same alkali. Different
from the case of a scalar condensate, we need three scattering
lengths to characterize the $s$-wave scattering between alike and
unlike atoms, which is referred to as intra-species and
inter-species scattering, respectively. The interplay between the
inter-species and intra-species scattering has a direct
consequence on the properties of the condensates, e.g. the density
profile \cite{ho,bigelow} and collective excitations
\cite{bigelow2}. For example, as shown in a pioneering work by Ho
and Shenoy \cite{ho}, the phase of trapped condensates can
smoothly evolve from interpenetrating to separate by changing the
atom numbers and/or the scattering lengths.

In addition to studying the stationary properties of BECs, the
coherent dynamics of a two species BEC can be investigated by
propagating coupled Gross-Pitaevskii equations. Previous work has
shown various interesting dynamics for two-species BECs
\cite{zhang,liu1,liu2,ricardo2}. For this paper, we focus on the
case of the two species being two hyperfine states of the same
alkali. One can manipulate such a two species BEC either by
turning on the population transfer between the two internal states
using a coupling microwave field
\cite{savage,hollanda,kuang,liu,you,xie} or by placing it in a
double well to allow for Josephson tunnelling
\cite{law,liang,yang,weiss,clark}. For the latter case, the double
well is fundamental to the study of the Josephson effect in BECs.
Smerzi and co-workers have shown that a scalar BEC trapped in a
double well exhibits rich dynamics such as Rabi oscillations and
macroscopic quantum self-trapping \cite{smerzi,smerzi2,smerzi3}.
It has also been shown that there exists non-Abelian Josephson
effect between two $F=2$ spinor BECs in double optical traps
\cite{liu3}. Therefore it is of great interest to study the
behavior of a two species BEC in a double well. Previous work on a
two species BEC has focused either on the population transfer or
on Josephson tunnelling. To the best of our knowledge, the
combination of both of them has not been addressed in the
literature. It is the purpose of this paper to discuss some
interesting phenomena involving both population transfer and
Josephson tunnelling.

%For instance, in a two species BEC with two hyperfine states of
%the same alkali, one can turn on the population transfer by
%coupling the two species with a microwave field and study the
%population dynamics \cite{savage,kuang,liu}. It has been shown
%that, in a two species BEC, collapses and revivals (CR) can also
%emerge \cite{kuang} just like in a scalar BEC \cite{wall}.

This paper is organized as follows. In Sec. II, we first introduce
our model Hamiltonian and various parameters. In Sec. III, we then
discuss some interesting regimes and give our numerical results.
We summarize our findings in Sec. IV. Finally, an appendix is
given in Sec. V.

\section{model}

\begin{figure}[htb]
\includegraphics[width=2.75in,angle=270]{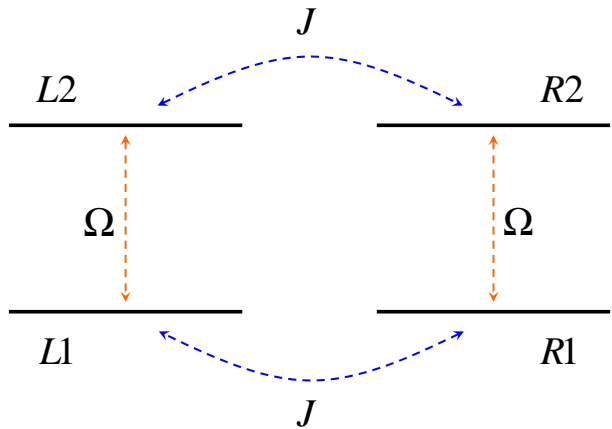}
\caption{(Color online) A schematic view of our model system where
the solid lines represent the four modes and the dashed lines
represents the coupling (red) and tunnelling (blue). The $s$-wave
scattering terms are not shown in this figure. } \label{figure1}
\end{figure}

In this paper, we consider a two species Bose-Einstein condensate
in a double well. We focus on the case where the two species are
for two different hyperfine spin states, e.g. $^{87}$Rb
condensates in $|1\rangle \equiv |F=1,m_F=-1\rangle$ and
$|2\rangle \equiv|F=2,m_F=1\rangle$, and they are subjected to the
same trapping potential. We adopt the well-known two-mode
approximation, i.e. $\Psi_{\alpha,j}({\bf r})\simeq a_{\alpha,j}
\psi_{\alpha}({\bf r})$. The Greek letter $\alpha=L(R)$ denotes
the left (right) well and the Roman letter $j=1(2)$ labels the two
species. $\psi_{\alpha}({\bf r})$ is the ground state solution of
the $\alpha$ well which is independent of the species.
$a_{\alpha,j}$ is the annihilation operator for species $i$ in the
$\alpha$ well, satisfying
$[a_{\alpha,j},a_{\beta,k}^\dagger]=\delta_{\alpha,\beta}
\delta_{j,k}$. Here $\delta_{p,q}$ is the Kronecker delta. In the
second quantization formalism, the Hamiltonian is ($\hbar=1$)
\begin{eqnarray}
H &=& \sum_{\alpha=L,R;j=1,2}\epsilon_{\alpha}
a_{\alpha,j}^\dagger a_{\alpha,j} + J \sum_{j} (a_{L,j}^\dagger
a_{R,j} +a_{L,j} a_{R,j}^\dagger )\nonumber\\
&+& \sum_{\alpha} \left[ {g_{11}^{(\alpha)} \over 2}
a_{\alpha,1}^\dagger a_{\alpha,1}^\dagger a_{\alpha,1}
a_{\alpha,1} + {g_{22}^{(\alpha)} \over 2} a_{\alpha,2}^\dagger
a_{\alpha,2}^\dagger a_{\alpha,2} a_{\alpha,2}  \right.\nonumber\\
&+& \left. g_{12}^{(\alpha)} a_{\alpha,1}^\dagger
a_{\alpha,2}^\dagger a_{\alpha,1} a_{\alpha,2} + \Omega
(a_{\alpha,1}^\dagger a_{\alpha,2}+a_{\alpha,1}
a_{\alpha,2}^\dagger) \right]  \label{fullH} .
\end{eqnarray}
$\epsilon_{\alpha}$ is the single particle energy in the $\alpha$
well. $2J= \int d{\bf r}\psi_L^*({\bf r})(T+V({\bf r}))\psi_R({\bf
r}) + \int d{\bf r}\psi_L({\bf r})(T+V({\bf r}))\psi_R^*({\bf r})$
is the intra-species Josephson tunnelling rate between the left
and right well which is referred to as tunnelling in the
following. Here $T$ and $V$ are the kinetic and potential (double
well) operator of a single particle, respectively. $J$ can be
tuned by changing the shape of the double well.
$g_{ij}^{(\alpha)}\equiv (4\pi \hbar^2 a_{ij}/m) \int d{\bf r}
|\psi_\alpha({\bf r})|^4$ is the interaction strength between
species $i$ and $j$ in the $\alpha$ well. $a_{ij}$ is the $s$-wave
scattering length between species $i$ and $j$. In this study, we
only consider the case of repulsive interactions, i.e. $a_{ij}>0$.
$m$ is the atom mass. $2\Omega$ is the Rabi frequency for the
inter-species population transfer which is referred to as coupling
in the following. It is proportional to the amplitude of the
microwave field which resonantly couples the two internal states.
For simplicity, we assume $J\ge 0$ and $\Omega\ge 0$. In addition,
we consider a symmetric double well in this paper, thus
$\epsilon_L=\epsilon_R$ and $\psi_L({\bf r})=\psi_R({\bf r})$.
Correspondingly, $g_{ij}^{(L)}=g_{ij}^{(R)}$ and the superscript
is then dropped afterwards. A schematic view of our model system
is shown in Fig. \ref{figure1}. We note that the Hamiltonian of
Eq. (\ref{fullH}) resembles the cross phase modulation between two
modes of light in a nonlinear Kerr medium. The terms involving
$\Omega$ and $J$ are analogous to the Hamiltonian of a beam
splitter. In the point view of quantum optics, our Hamiltonian
describes the combination of nonlinear beam splitters.

We assume that initially all $N$ atoms are of species 1 and
localized in the left well, i.e. $|\psi(t=0)\rangle\propto
(a_{L,1}^\dagger)^N|\rm vac\rangle$. The coupling $\Omega$ and $J$
are turned on at $t=0$ and we investigate the subsequent dynamics.

\section{various regimes}

We categorize our model into three regimes: the strong
coupling/tunnelling regime $(\Omega,J) \gg g_{ij}(N-1)$, the
strong interaction regime $g_{ij}(N-1) \gg (\Omega,J)$, and the
intermediate regime where the terms of $g_{ij}(N-1)$, $\Omega$,
and $J$ have equally important contributions to the dynamics. In
the following, we will discuss these three regimes separately.

We first consider the strong coupling/tunnelling regime so the
collisional interaction terms are neglected. The Hamiltonian is
\begin{eqnarray}
H_C^{(o)} &=& \Omega (a_{L,1}^\dagger a_{L,2}+a_{R,1}^\dagger
a_{R,2}+a_{L,1}
a_{L,2}^\dagger+a_{R,1} a_{R,2}^\dagger)\nonumber\\
 &+& J(a_{L,1}^\dagger a_{R,1}+a_{L,2}^\dagger a_{R,2}+a_{L,1}
a_{R,1}^\dagger+a_{L,2} a_{R,2}^\dagger). \nonumber\\
\end{eqnarray}
It is easy to verify that the terms in the bracket involving
$\Omega$ commute with those involving $J$. $H_C^{(o)}$ can be
written in a more compact form $H_C^{(o)}={\bf a}^\dagger \cdot
{\bf M} \cdot {\bf a}$ where ${\bf a}=(a_{L,1},a_{L,2},
a_{R,1},a_{R,2})^{\rm T}$. The matrix ${\bf M}$ depends only on
$\Omega$ and $J$. This Hamiltonian is quadratic, so it can be
simplified in a standard manner by diagonalizing ${\bf M}$.
Explicitly, by the following transformation,
\begin{eqnarray}
f_1 &=& {1\over 2} ( a_{L,1}+a_{L,2}+a_{R,1}+a_{R,2}), \nonumber\\
f_2 &=& {1\over 2} ( a_{L,1}-a_{L,2}-a_{R,1}+a_{R,2}), \nonumber\\
g_1 &=& {1\over 2} ( a_{L,1}+a_{L,2}-a_{R,1}-a_{R,2}), \nonumber\\
g_2 &=& {1\over 2} ( a_{L,1}-a_{L,2}+a_{R,1}-a_{R,2}).
\end{eqnarray}
$[f_j,f_k^\dagger]=\delta_{jk}$ and
$[g_j,g_k^\dagger]=\delta_{jk}$ with all other commutators being
zero. The conservation of atom numbers leads to $f_1^\dagger
f_1+f_2^\dagger f_2+g_1^\dagger g_1+g_2^\dagger g_2=N$. The
Hamiltonian can be rewritten in the form of new operators
\begin{eqnarray}
H_C=(\Omega+J)(f_1^\dagger f_1-f_2^\dagger f_2) + (\Omega-J)
(g_1^\dagger g_1-g_2^\dagger g_2),
\end{eqnarray}
where constant terms have been dropped out. We note that the net
effect of coupling and tunnelling can be either constructive
($\Omega+J$) where the coupling and tunnelling are in phase or
destructive ($\Omega-J$) where they are out of phase, analogous to
the center of mass and the relative motion of two particles,
respectively. The corresponding initial condition is
$|\psi(t=0)\rangle\propto
[f_1^\dagger+f_2^\dagger+g_1^\dagger+g_2^\dagger]^N|\rm
vac\rangle$.

The wave function at any time $t$ can be obtained with the help of
the following identity
\begin{eqnarray}
&& e^{-iH_C t}(f_1^\dagger+f_2^\dagger+g_1^\dagger+g_2^\dagger)e^{iH_C t}\nonumber\\
&=&e^{-i(\Omega+J)t} f_1^\dagger+e^{i(\Omega+J)t}
f_2^\dagger+e^{-i(\Omega-J)t} g_1^\dagger+e^{i(\Omega-J)t}
g_2^\dagger \nonumber.
\end{eqnarray}
Therefore,
\begin{eqnarray}
&& |\psi(t)\rangle= {1\over 2^N\sqrt{N!}} \left[ e^{-i(\Omega+J)t}
f_1^\dagger +e^{i(\Omega+J)t} f_2^\dagger \right.\nonumber\\
&& \left.+e^{-i(\Omega-J)t} g_1^\dagger+e^{i(\Omega-J)t}
g_2^\dagger   \right]^N |\rm vac\rangle . \nonumber
\end{eqnarray}
Substituting $a_{\alpha,j}^\dagger$ back into the above
expression, we obtain the normalized wave function
\begin{eqnarray}
|\psi(t)\rangle&=& {1\over \sqrt{N!}}
\left[\sum_{\alpha=L,R;j=1,2} S_{\alpha,j}(t) a_{\alpha,j}^\dagger
\right]^N |\rm vac\rangle,  \nonumber\\
S_{L,1}(t)&=&{ {\rm cos}(\Omega+J)t+{\rm cos}(\Omega-J)t \over 2}, \nonumber\\
S_{R,2}(t)&=&{ {\rm cos}(\Omega+J)t-{\rm cos}(\Omega-J)t \over 2}, \nonumber\\
S_{L,2}(t)&=&-i{ {\rm sin}(\Omega+J)t+{\rm sin}(\Omega-J)t \over 2}, \nonumber\\
S_{R,1}(t)&=&-i{ {\rm sin}(\Omega+J)t-{\rm sin}(\Omega-J)t \over
2}. \label{wf}
\end{eqnarray}

Various quantities can be calculated directly from the wave
function. For instance, the first order coherence between modes
$a_{\alpha,j}$ and $a_{\beta,k}$ is found to be
\begin{eqnarray}
g^{(1)}_{\alpha,j;\beta,k}(t)&=&{1\over N}\langle
\psi(t)|a_{\alpha,j}^\dagger a_{\beta,k}|\psi(t) \rangle \nonumber\\
&=&  S^*_{\alpha,j}(t) S_{\beta,k}(t).
\end{eqnarray}
Especially, the fraction of species $j$ in the $\alpha$ well is
given by
\begin{eqnarray}
n_{\alpha,j}(t) =g^{(1)}_{\alpha,j;\alpha,j}(t)
=|S_{\alpha,j}(t)|^2.
\end{eqnarray}
Higher order coherence terms can be derived straightforwardly with
similar patterns and will not be presented here. We note that at
any time $t>0$, there is a balancing condition
$n_{L1}/n_{L2}=n_{R1}/n_{R2}$ for the given initial state.

\begin{figure}[htb]
\includegraphics[width=3.in]{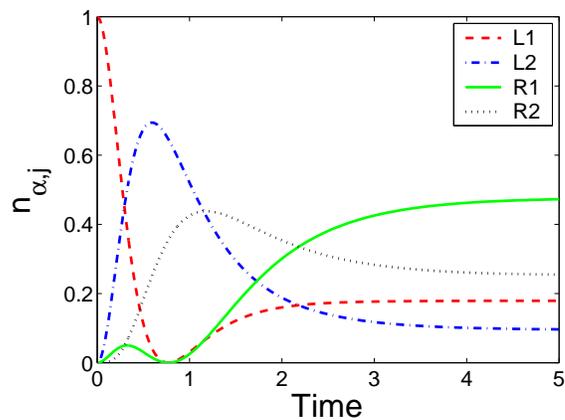}
\caption{(Color online) Fraction of each component as a function
of time. $\Omega=3.2,J=1.3,\tau=\pi/4, g_{11}=g_{22}=g_{12}=0$.}
\label{figure2}
\end{figure}

Our conclusion also applies to the case with temporal modulations
of $\Omega$ and $J$ if we make the replacement
\begin{eqnarray}
e^{-iH_C t}\to e^{-i\int_0^t  H_C (t')dt'}.
\end{eqnarray}
The time ordered exponential is not needed here since the
hamiltonian commutes at any time in this representation.
Therefore, the effect of time dependence is determined from two
pulse areas, i.e. $\int_0^t (\Omega(t')+J(t')) dt'$ and $\int_0^t
(\Omega(t')-J(t')) dt'$. If we assume a form of
$\Omega(t')=\Omega_0 e^{-t'/\tau}$ and $J=J_0 e^{-t'/\tau}$, then
the two pulse areas are $(\Omega_0+J_0)\tau (1-e^{-t/\tau})$ and
$(\Omega_0-J_0)\tau (1-e^{-t/\tau})$ for the case of in phase and
out of phase, respectively. For a long enough time $t\gg \tau$,
the exponential terms in the above pulse areas can be neglected
and the tunnelling dynamics are suppressed. The wave function
takes the same form as Eq. \ref{wf} except for the replacements
$\Omega\to \Omega_0, J\to J_0,$ and $t\to \tau$. A numerical
example is shown in Fig. \ref{figure2}. Unless otherwise
specified, the energy and time scale are $\omega_0$ and
$1/\omega_0$, respectively, where $\omega_0$ is the angular
frequency obtained by approximating one well as harmonic.

\begin{figure}[htb]
\includegraphics[width=3.in]{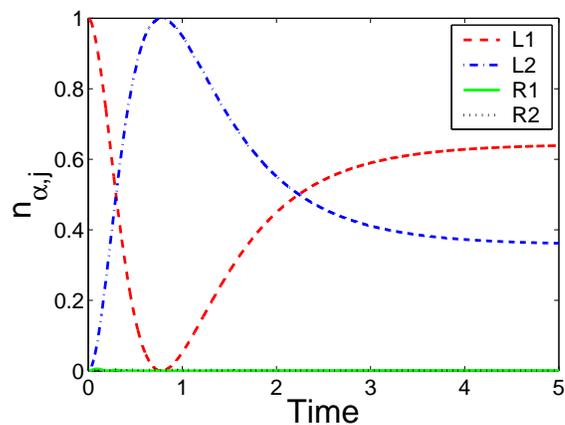}
\caption{(Color online) Fraction of each component as a function
of time. $\Omega=3.2,J=1.3,\tau=\pi/4, g_{11}=g_{22}=g_{12}=35$.}
\label{figure3}
\end{figure}

In the opposite regime where the nonlinear interaction dominates,
the coupling between the two species is still on resonance.
However, the tunnelling between the left and right well is frozen
due to the large detuning. For instance, the energy difference
between the two Fock states
$|N\rangle_{L1}|0\rangle_{L2}|0\rangle_{R1}|0\rangle_{R2}$ and
$|N-1\rangle_{L1}|0\rangle_{L2}|1\rangle_{R1}|0\rangle_{R2}$ is
$\propto g_{11}(N-1)\gg J$ for large $N$. So the dynamics are
restricted to the left well which is essentially a two-level
system. A numerical example is shown in Fig. \ref{figure3} where
we can clearly see that the dynamics are restricted to the two
species in the left well and the tunnelling to the right well is
suppressed.

In the intermediate regime, we cannot obtain solutions in closed
form in general. In the following, we will consider a specific
case where we can obtain some analytical results. We assume
$g_{11}=g_{22}=0$ and the term involving $g_{12}$ is a
perturbation to the coupling/tunnelling. We denote $g_{12}=g$ in
the following. Motivated by the strong coupling/tunnelling case,
we adopt the following transformation
\begin{eqnarray}
a_{L,1}={1\over 2} [e^{i\omega_+ t} f_1 + e^{-i\omega_+ t}
f_2 + e^{i\omega_- t} g_1 +e^{-i\omega_- t} g_2 ],\nonumber\\
a_{L,2}={1\over 2} [e^{i\omega_+ t} f_1 - e^{-i\omega_+ t}
f_2 +e^{i\omega_- t} g_1 -e^{-i\omega_- t} g_2] ,\nonumber\\
a_{R,1}={1\over 2} [e^{i\omega_+ t} f_1 - e^{-i\omega_+ t}
f_2 -e^{i\omega_- t} g_1 +e^{-i\omega_- t} g_2] ,\nonumber\\
a_{R,2}={1\over 2} [e^{i\omega_+ t} f_1 + e^{-i\omega_+ t} f_2 -
e^{i\omega_- t} g_1 -e^{-i\omega_- t} g_2], \label{newT}
\end{eqnarray}
where $\omega_{\pm}\equiv 2(\Omega\pm J)$. $f_j$ and $g_j$ obey
the same commutation conditions as in the strong
coupling/tunnelling case. Substituting $a_{\alpha,j}$ into the
Hamiltonian (\ref{fullH}), we then invoke the rotating wave
approximation (RWA) to eliminate the fast oscillating terms and
only keep resonant terms. Explicitly, RWA requires $g\ll
(\Omega,J,|\Omega-J|,|2\Omega-J|,|\Omega-2J|)$. The new
Hamiltonian becomes
\begin{eqnarray}
 H_R &=&{g\over 8}\left(f_1^\dagger f_1^\dagger f_1 f_1 +
 f_2^\dagger f_2^\dagger f_2 f_2
+ g_1^\dagger g_1^\dagger g_1 g_1 \right. \nonumber\\
&&\left. + g_2^\dagger g_2^\dagger g_2 g_2 + 4 f_1^\dagger
g_1^\dagger f_1 g_1 + 4 f_2^\dagger g_2^\dagger f_2
g_2 \right)\nonumber\\
&& +(\Omega+J)(f_1^\dagger f_1-f_2^\dagger f_2) \nonumber\\
&& + (\Omega-J) (g_1^\dagger g_1-g_2^\dagger g_2). \label{HR}
\end{eqnarray}
This Hamiltonian is separable in two groups of modes $(f_1,g_1)$
and $(f_2,g_2)$. Each group of modes, $f_j$ and $g_j$, are coupled
by the mean field interaction and the detuning between them is
$2J$. The above Hamiltonian is already diagonal in the basis of
$f_j$ and $g_j$, so we label the Fock eigenstate as $|mnpq)$ in
the basis order of $f_1, f_2, g_1, g_2$. The corresponding energy
is $E_{mnpq} = E_{m,p}+E_{n,q} $ with
\begin{eqnarray}
E_{m,p} &=& {g\over 8}(m^2 -m +p^2-p+4m p) \nonumber\\
&& +(\Omega+J)m +(\Omega-J)p  ,\nonumber\\
E_{n,q} &=& {g\over 8}(n^2-n+q^2-q+4nq) \nonumber\\
&& -(\Omega+J)n -(\Omega-J)q  .\nonumber
\end{eqnarray}

%For instance,
%\begin{eqnarray}
%&& a_{L,1}^\dagger a_{L,2}^\dagger a_{L,1}a_{L,2} \to {1\over
%16}\left[ f_1^\dagger f_1^\dagger f_1 f_1 + f_2^\dagger
%f_2^\dagger f_2 f_2 \right. \nonumber\\
%&&\left. + g_1^\dagger g_1^\dagger g_1 g_1 + g_2^\dagger
%g_2^\dagger g_2 g_2 + 4 f_1^\dagger g_1^\dagger f_1 g_1 +4
%f_2^\dagger g_2^\dagger f_2 g_2 \right]. \nonumber
%\end{eqnarray}

Note that as $|\rm vac\rangle=|\rm vac)$, the wave function can be
computed in the new basis as
\begin{eqnarray}
&& |\psi(t)\rangle \nonumber\\
&=&  e^{-iH_R t} {1\over 2^N \sqrt{N!}}
(f_1^\dagger+f_2^\dagger+g_1^\dagger+g_2^\dagger)^N|\rm vac\rangle
\nonumber\\
&=&  e^{-iH_R t} {1\over 2^N } \sum_{mnpq} \delta_{N,m+n+p+q}
{\sqrt{N!}\over \sqrt{m!n!p!q!}} |mnpq) \nonumber\\
&=& {1\over 2^N } \sum_{mnpq} \delta_{N,m+n+p+q} {\sqrt{N!}\over
\sqrt{m!n!p!q!}} e^{-i E_{mnpq}t} |mnpq).  \nonumber\\
\end{eqnarray}

The fraction of each species $n_{\alpha,j}$ can be computed with
the help of the following results where the average is taken with
respect to the wave function $|\psi(t)\rangle$:
\begin{eqnarray} \langle f_1^\dagger f_1 \rangle &=& \langle
f_2^\dagger f_2 \rangle = \langle g_1^\dagger g_1 \rangle =
\langle g_2^\dagger g_2 \rangle = {N\over 4},  \nonumber\\
\langle f_1^\dagger f_2 \rangle &=& e^{2i(\Omega+J)t} {N\over 4}
\left({\cos (tg/4) +\cos (tg/2) \over 2} \right)^{N-1},  \nonumber\\
%%%%%%%%%%%%%%%%
\langle f_1^\dagger g_1 \rangle &=& e^{2iJt} {N\over 4}
\left({1+\cos (tg/4) \over 2} \right)^{N-1},  \nonumber\\
%%%%%%%%%%%%%%%%
\langle f_1^\dagger g_2 \rangle &=&  e^{2i\Omega t} {N\over 4}
\left({\cos (tg/4)+\cos (tg/2) \over 2} \right)^{N-1},  \nonumber\\
%%%%%%%%%%%%%%%%
\langle f_2^\dagger g_1 \rangle &=&  e^{-2i\Omega t} {N\over 4}
\left({\cos (tg/4)+\cos (tg/2) \over 2} \right)^{N-1},  \nonumber\\
%%%%%%%%%%%%%%%%
\langle f_2^\dagger g_2 \rangle &=& e^{-2iJ t} {N\over 4}
\left({1+\cos (tg/4) \over 2} \right)^{N-1},  \nonumber\\
%%%%%%%%%%%%%%%%
\langle g_1^\dagger g_2 \rangle &=& e^{2i(\Omega-J)t} {N\over 4}
\left({\cos (tg/4)+\cos (tg/2) \over 2} \right)^{N-1}. \nonumber\\
\end{eqnarray}
In the Appendix, we give a derivation of $\langle f_1^\dagger f_2
\rangle$. Other average values can be derived along similar lines.

After some algebra, we obtain
\begin{eqnarray}
&& n_{L,1} = {1\over 4} \left[ 1+ \cos (2Jt) \left({1+\cos (tg/4)
\over 2} \right)^{N-1}  \right. \nonumber\\
&& \left. + \cos (2\Omega t) (1+ \cos (2Jt) ) \left({\cos
(tg/4)+\cos (tg/2) \over 2} \right)^{N-1}  \right], \nonumber\\
\label{nL1}
\end{eqnarray}
%%%%%%%%%%%%%%%%%
\begin{eqnarray}
&& n_{L,2} = {1\over 4} \left[ 1+ \cos (2Jt) \left({1+\cos
(tg/4) \over 2}\right)^{N-1}  \right.\nonumber\\
&& \left. -  \cos (2\Omega t) (1+ \cos (2Jt) ) \left({\cos
(tg/4)+\cos (tg/2) \over 2}  \right)^{N-1} \right], \nonumber\\
\label{nL2}
%%%%%%%%%%%%%%%%%
\end{eqnarray}
\begin{eqnarray}
&& n_{R,1} = {1\over 4} \left[ 1  - \cos (2Jt) \left({1+\cos
(tg/4)
\over 2} \right)^{N-1}  \right. \nonumber\\
&& \left. +  \cos (2\Omega t) ( 1- \cos (2Jt) ) \left({\cos
(tg/4)+\cos (tg/2) \over 2} \right)^{N-1}  \right], \nonumber\\
\label{nR1}
%%%%%%%%%%%%%%%%%
\end{eqnarray}
\begin{eqnarray}
&& n_{R,2} = {1\over 4} \left[ 1- \cos (2Jt) \left({1+\cos
((tg/4))
\over 2} \right)^{N-1}  \right. \nonumber\\
&& \left.  -  \cos (2\Omega t) (1- \cos (2Jt) ) \left({\cos
(tg/4)+\cos (tg/2) \over 2} \right)^{N-1}\right]. \nonumber\\
\label{nL2}
\end{eqnarray}
Therefore, in the case of the vanishing intra-species scattering,
$n_{\alpha,j}$ will be subjected to collapses and revivals (CR).
CR is a quantum mechanical effect which is well known in quantum
optics. It is also found for a scalar BEC \cite{wall} as well as
for a two-species condensate in a single well \cite{kuang}. The
nonlinearity $g$ determines the envelope of the revivals as well
as the time separation between the adjacent collapse and revival.
It is easy to show that the width of the revival in the time
domain is $\propto 1/(g\sqrt{N-1})$ for $N>1$ and the separation
between a neighboring CR is $8\pi/g$. The coupling $\Omega$ and
tunnelling $J$ determine the detailed structures of the
oscillation inside the revival envelopes. Here we want to point
out a major difference between our study and that of a two species
BEC in a single well. In our case, CR is not observed for general
scattering lengths $a_{ij}$. When the intra-species scattering
lengths are not negligible, the population dynamics usually
display quite complicated temporal patterns. We note that, for
large $N$, since $n_{\alpha,j}$ are all small except for
$gt/4=2n\pi$, we have
\begin{eqnarray}
&& {1+\cos (tg/4) \over 2} \sim e^{-\sin^2 (tg/8)}, \nonumber\\
&& {\cos (tg/4)+\cos (tg/2) \over 2}\sim e^{-5\sin^2 (tg/8)}.
\end{eqnarray}
So up to some factors, our expressions take the same exponential
form as those in Ref. \cite{kuang}, e.g. Eq. (31). This is
expected because the {\it relative} difference between a coherent
state and a Fock state is small for large (average) atom numbers.
We also note that, for $N=1$, $n_{\alpha,j}$ reduces to the
results of the strong coupling/tunnelling regime.

The population difference between the two wells/species takes a
relatively simpler form which is shown below
\begin{eqnarray}
D_{LR} &\equiv & (n_{L,1}+n_{L,2})-(n_{R,1}+n_{R,1}) \nonumber\\
&=& \cos (2Jt) \left({1+\cos (tg/4) \over 2} \right)^{N-1}, \nonumber\\
D_{12} &\equiv & (n_{L,1}+n_{R,1})-(n_{L,2}+n_{R,2}) \nonumber\\
&=& \cos (2\Omega t) \left({\cos (tg/4)+\cos (tg/2) \over 2}
\right)^{N-1} .
\end{eqnarray}
Again, both of them show CR which are modulated by a sinusoidal
oscillation with individual frequency $2J$ or $2\Omega$, i.e. $J$
($\Omega$) has no effect on $D_{12}$ ($D_{LR}$). This is not a
surprising result even in the presence of nonlinearity since we
are dealing with a symmetric double well.

\begin{figure}[htb]
\includegraphics[width=3.in]{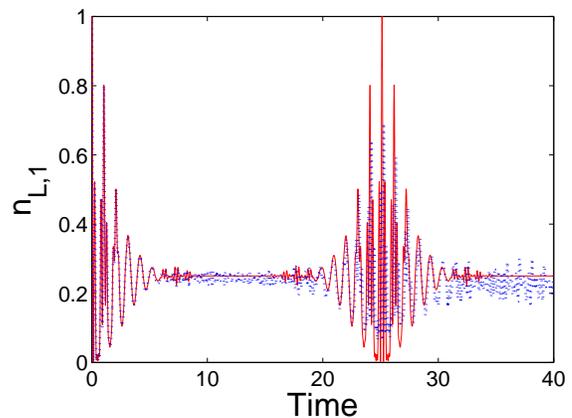}
\caption{(Color online) $n_{L,1}$ as a function of time. The red
solid curve is for result Eq. (\ref{nL1}). The blue dotted curve
is for exact numerical simulations with the full Hamiltonian of
Eq. (\ref{fullH}). $\Omega=2.4,J=0.6,
g_{11}=g_{22}=0,g_{12}=g=0.2$, and $N=6$. Time is in units of
$1/g$.} \label{figure4}
\end{figure}

A numerical example of $n_{L,1}$ as function of time is shown in
Fig. \ref{figure4} where the (red) solid curve is for result of
Eq. (\ref{nL1}) and the (blue) dotted curve is for exact numerical
simulations with the full Hamiltonian of Eq. (\ref{fullH}). We can
see that RWA does capture the essential features of the dynamics.
It reproduces the results very accurately for the initial time.
However, it is only qualitatively accurate for long times, e.g.
RWA overestimates the peaks of population oscillation for the
first revival. Both RWA and exact numerical simulations predict
the CR phenomenon. Therefore, CR for a two species BEC is not an
artifact of RWA, but rather a consequence of phase coherence under
nonlinear interaction.

\begin{figure}[htb]
\includegraphics[width=3.in]{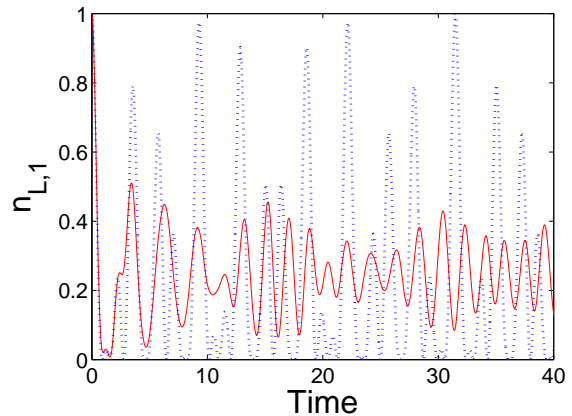}
\caption{$n_{L,1}$ as a function of time using exact numerical
simulations with the full Hamiltonian of Eq. (\ref{fullH}).
$\Omega=1,J=1.7$, $N=6$. The red solid curve is for $g11=g22=0.45$
and $g12=0.3$. The blue dotted curve is for the noninteracting
case $g11=g22=g12=0$. } \label{figure5}
\end{figure}

The general case can be attacked numerically. In Fig.
\ref{figure5}, we show our numerical results for $n_{L,1}$ as
function of time. The parameters are chosen so as to fall into the
intermediate regime. For comparison, we also show the results of
the strong coupling-tunneling case. We can see that, with
appropriate nonlinearity, the dynamics of $n_{L,1}$ exhibit
completely different and more complicated patterns than the strong
coupling-tunneling case. The nonlinearity not only reduces the
oscillation amplitude but also destroys the periodicity
\cite{note1}. This reduction in the amplitude is due to the
effective damping which is a common feature in interacting
condensates \cite{sun1,sun2}.

Now we briefly discuss a possible experimental implementation of
our proposal. Our model can be conveniently realized in an optical
super-lattice by superimposing two-color lasers
\cite{phil,bloch1}. Each lattice site consists of a double well
potential. $J$ is determined from the shape of the double well
which can be tuned by engineering the ratio of the two laser
intensities. The $s$-wave scattering $g_{ij}$ can be tuned by
using the well established Feshbach resonance. $\Omega$ can be
tuned by changing the intensities of two microwave/RF fields both
of which are coupled to the third intermediate level with a large
detuning, which can be adiabatically eliminated. The relative
strength in our model Hamiltonian can thus be tuned over a wide
range. For the CR as shown in Fig. \ref{figure4}, we give an order
of magnitude estimate of typical experimental parameters here. If
we approximate each lattice site as a harmonic trap with trapping
frequency $\omega\sim 10$kHz. The intra-species interaction
strength $g$ can be tuned to be $\sim 100$Hz with the
corresponding $a_{12}\simeq 25 a_B$ ($a_B$ is the Bohr radius).
The Josephson tunnelling rate $J$ and the coupling strength
$\Omega$ is $300$Hz and $1.2$kHz, respectively, both of which can
be implemented using the state-of-the-art technology.

\section{conclusion}

To conclude, we have studied the dynamics of a two-species
Bose-Einstein condensate in a double well. Such a system is
characterized by the following three factors. (1) The $s$-wave
scattering $-$ $a_{11}$ and $a_{22}$ for intra-species scattering
and $a_{12}$ for inter-species scattering; (2) Josephson
tunnelling between the two wells; (3) population transfer between
the two species driven by a resonant microwave field. We discuss
the dynamics for three interesting regimes where we can obtain
analytical results. (a) The strong coupling /tunnelling regime
where the nonlinearity is negligible. We find the net effect of
the coupling/tunnelling is either constructive or destructive; (b)
the strong nonlinearity regime where Josephson tunnelling is
suppressed and the system behaves like a simple two-level system;
(c) the intermediate regime. For this case we only consider a
specific example of vanishing intra-species scattering and weak
inter-species scattering. We find collapses and revivals in the
population dynamics. For the general case, we attack this problem
numerically. We find the dynamics is rather complicated. The
nonlinearity forces the system out of periodicity. We hope our
work can be helpful to the study of quantum coherence of
two-species BECs.

We deeply appreciate Dr. F. Robicheaux at Auburn University and
Dr. L. You at Georgia Institute of Technology for enlightening
discussions and helpful comments on the manuscript. We also thank
J. Ludlow for proof reading our manuscript. This work is supported
by grants with NSF.

\section{appendix}
In this appendix, we show the procedure to derive $\langle
f_1^\dagger f_2 \rangle$. Other average values can be derived
along similar lines.
\begin{eqnarray}
%%%%%%%%%%%%%%%%
\langle f_1^\dagger f_2 \rangle &=&  {1\over 4^N } \sum_{mnpq}
\delta_{N,m+n+p+q} {N!\over m!(n-1)!p!q!}  \nonumber\\
&&  e^{i(E_{m+1,p}+E_{n-1,q}-E_{m,p}-E_{n,q})t} \nonumber\\
&=& {N\over 4^N } \sum_{mnpq}
\delta_{N-1,m+n-1+p+q} {(N-1)!\over m!(n-1)!p!q!}  \nonumber\\
&&  e^{i(E_{m+1,p}+E_{n-1,q}-E_{m,p}-E_{n,q})t} \nonumber\\
&=& {N\over 4^N } \sum_{mnpq}
\delta_{N-1,m+n-1+p+q} {(N-1)!\over m!(n-1)!p!q!}  \nonumber\\
&& e^{it(2(\Omega+J)+g/4)} e^{it{g\over 4} ( m - n + 2 p - 2 q)} \nonumber\\
&=& e^{it 2(\Omega+J)} {N\over 4^N } \sum_{mnpq} {1\over
2\pi} \int_0^{2\pi} e^{i(m+n-1+p+q-(N-1))\phi} d\phi  \nonumber\\
&&   {(N-1)!\over m!(n-1)!p!q!} e^{it{g\over 4} ( m - (n-1) + 2 p - 2 q)} \nonumber\\
&=& e^{it 2(\Omega+J)} {N\over 4^N }(N-1)! {1\over 2\pi}
\int_0^{2\pi} d\phi e^{-i(N-1)\phi} \nonumber\\
&&  \sum_{mnpq} e^{im\phi} { e^{it m g/4}\over m!} e^{i(n-1)\phi}
{ e^{-it (n-1) g/4}\over (n-1)!} \nonumber\\
&& e^{ip\phi} { e^{it p g/2}\over p!} e^{iq\phi} { e^{-it q
g/2}\over q!} \nonumber\\
&=& e^{it 2(\Omega+J)} {N\over 4^N }(N-1)! {1\over 2\pi}
\int_0^{2\pi} d\phi e^{-i(N-1)\phi} \nonumber\\
&& {\rm exp}[e^{i\phi+i(tg/4)}] {\rm exp}[e^{i\phi-i(tg/4)}] \nonumber\\
&& {\rm exp}[e^{i\phi+i(tg/2)}] {\rm exp}[e^{i\phi-i(tg/2)}] \nonumber\\
&=& e^{it 2(\Omega+J)} {N\over 4^N }(N-1)! {1\over 2\pi}
\int_0^{2\pi} d\phi e^{-i(N-1)\phi} \nonumber\\
&& \exp[e^{i\phi} 2(\cos (tg/4)+\cos (tg/2))]  \nonumber\\
&=& e^{it 2(\Omega+J)} {N\over 4^N }(N-1)! {1\over 2\pi}
\int_0^{2\pi} d\phi e^{-i(N-1)\phi} \nonumber\\
&& \sum_k {1\over k!} e^{ik\phi} 2^k (\cos (tg/4)+\cos (tg/2))^k  \nonumber\\
&=& e^{it 2(\Omega+J)} {N\over 4^N }(N-1)! \nonumber\\
&& {1\over (N-1)!} 2^{N-1} (\cos (tg/4)+\cos (tg/2))^{N-1} \nonumber\\
&=& e^{it 2(\Omega+J)} {N\over 4} \left({\cos (tg/4) +\cos (tg/2)
\over 2} \right)^{N-1}  \nonumber
\end{eqnarray}


\begin{thebibliography}{99}
\bibitem{note1}
In the strong coupling/tunnelling regime, the exact periodicity
only exists for commensurable $\Omega$ and $J$. Assuming $m$ and
$n$ are the smallest integers satisfying $\Omega/J=m/n$, the
period is $T=\min(m,n)\pi$ where $\min(m,n)$ denotes the minimum
of $m$ and $n$.
\end{thebibliography}
\end{document}